\documentclass{llncs}

\usepackage{cite}
\usepackage{amsmath,amssymb,amsfonts}
\usepackage{graphicx}
\usepackage{textcomp}
\usepackage{xcolor}
\usepackage{bm}
\usepackage[caption=false, font=footnotesize]{subfig}
\usepackage{booktabs}
\usepackage{tabularx}
\usepackage{colortbl}
\usepackage{xcolor}

\definecolor{webred}{rgb}{0.5,0,0}
\definecolor{webblue}{rgb}{0,0,0.8}

\usepackage{url}

\usepackage{algorithm}
\usepackage[noend]{algpseudocode}
\usepackage{bm}

\newcommand{\func}[1]{\textsf{#1}}
\newcommand{\var}[1]{\textbf{#1}}

\newcommand{\norm}[1]{\left\lVert#1\right\rVert}

\renewcommand{\int}[1]{\overline{\bm{#1}}}

\title{Efficient $n$-to-$n$ Collision Detection \\ for Space Debris using 4D AABB Trees (Extended Report)}

  \author{
    Stanley Bak\inst{1}
    \and
    Kerianne Hobbs\inst{2}
}

  \institute{
     Safe Sky Analytics \\
   Manlius, NY, USA \\
   \email{stanleybak@gmail.com} \\
   ~
   \and
  Air Force Research Laboratory,
  Dayton, OH, USA\\
  \email{kerianne.hobbs@us.af.mil} \\
  Georgia Institute of Technology \\
  \email{kerianne@gatech.edu} \\
 }

\begin{document}

\maketitle

\begin{abstract}
Collision detection algorithms are used in aerospace, swarm robotics, automotive, video gaming, dynamics simulation and other domains.
As many applications of collision detection run online, timing requirements are imposed on the algorithm runtime:
algorithms must, at a minimum, keep up with the passage of time.
In practice, this places a limit on the number of objects, $\bm{n}$, that can be tracked at the same time.
In this paper, we improve the scalability of collision detection, effectively raising the limit $\bm{n}$ for online object tracking.

The key to our approach is the use of a four-dimensional axis-aligned bounding box (AABB) tree,
which stores each object's three-dimensional occupancy region in space during a one-dimensional interval of time.
This improves efficiency by permitting per-object variable times steps.
Further, we describe partitioning strategies that can decompose the 4D AABB tree search into
several smaller-dimensional problems that can be solved in parallel.
We formalize the collision detection problem and prove our algorithm's correctness.
We demonstrate the feasibility of online collision detection for an orbital space debris application,
using publicly available data on the full catalog of $\bm{n}=16848$ objects provided by \url{www.space-track.org}.

\end{abstract}

\makeatletter
\def\blfootnote{\xdef\@thefnmark{}\@footnotetext}
\makeatother
\blfootnote{DISTRIBUTION A. Approved for public release; Distribution unlimited. Approval AFRL PA case number 88ABW-2018-4991, 05 OCT 2018.}

\section{Introduction}

The online prediction of collisions between large numbers of objects is important for many domains.
In the \$100 billion-per-year\footnote{\url{www.statista.com/statistics/246888/value-of-the-global-video-game-market}}
video game market, for example, collision detection must often be performed in real-time as the game is played.
In reality, the timing budget is a fraction of real-time since other operations, such as game logic and rendering, also take time.
Faster-than-real-time collision prediction is also essential for many cyber-physical systems with safety critical requirements,
where colliding physical objects could result in a significant financial or capability loss,
or even the loss of life.

The prediction problem becomes difficult when there a large number of objects moving within a space, and any object can
potentially collide with any other object (the $n$-to-$n$ problem).
Consider, for example, trying to predict collisions among \texttildelow5000 aircraft concurrently in the sky over the United States\footnote{\url{www.faa.gov/air\_traffic/by\_the\_numbers}},
or trying to find collisions among the \texttildelow23000 objects larger than 10 cm in orbit around Earth~\cite{Hildreth2014}.
Upcoming applications such as swarm robotics also demand collision-free maneuvering of large numbers of agents in tight areas.
In these environments, efficient collision prediction enables path deconfliction and collision avoidance.

Collision detection algorithms typically consist of two main phases, where a first phase (the broad phase)
uses overapproximations of objects to check for potential collisions, and a second phase (the narrow phase) performs
more expensive, exact analysis.
For example, the broad phase may use bounding boxes of objects and can quickly reject most potential collisions.
If the bounding boxes of objects intersect, the narrow phase may look at the exact geometry of objects and see if there was a true collision.
In this work, we focus on the scalability of the first, broad phase of collision detection.
Further, we focus on collision detection, not collision resolution, which is the system's reaction when a collision occurs.
%
%The problem, therefore, is to either find the first time and objects that collide, or to assure that no objects will collide up to a given time bound.

The main contributions of this paper are:

\begin{itemize}
\item We formalize the broad phase collision detection problem and provide an efficient and provably-correct solution using a 4D version of the AABB tree data structure~\cite{aabb}.
\item We propose static decomposition methods that leverage parallel processing to further increase scalability.
\item We demonstrate that the approach is capable of online space debris collision detection, with $n=16838$ objects on multiple platforms.
\end{itemize}

This paper is organized as follows.
We first present definitions and formally define the collision prediction problem in Section~\ref{sec:preliminaries}.
In Section~\ref{sec:collision_detection}, we review existing approaches based on brute force checking and existing AABB tree collision detection
methods.
The primary contribution of this work is the 4D AABB tree collision detection algorithm presented in Section~\ref{sec:AABB_4D}, with
the proof of its correctness presented in Section~\ref{sec:correctness}.
A discussion of domain-specific state space decomposition provided in Section~\ref{sec:band_splitting}.
Finally, the approach is evaluated on an orbital object collision detection application in Section~\ref{sec:space_debris}.
Related work is given in Section~\ref{sec:related}, followed by a conclusion.

\section{Preliminaries}
\label{sec:preliminaries}

There are many variants of the $n$-to-$n$ collision detection problem, so it is important precisely define the problem being solved.
Further, to clarify the presentation, correctness proof, and eventual evaluation, we consider a number of simplifications,
outlined in the next paragraph.
We expect, however, that the qualitative scalability results of our proposed algorithm to be similar in modified scenarios where some
of the assumptions may not hold.

Although we are interested in online performance, we define the problem in a static setting,
and will then later measure its performance for suitability in online application.
We assume there are a fixed number of objects, and we check for collisions up to a predefined time bound at multiples of a given time step.
The problem is to either find the first time and pair of objects that collide, or to ensure that no objects collide up to the time bound.
Since we are doing broad-phase collision detection, we will consider objects with a fixed infinity-norm (box) radius,
although extending to general boxes is straightforward.

Formally, the considered collision detection problem deals with objects that take up 3D space that change position over time.
\begin{definition}[\textbf{World Object}]
  A \emph{world object} or simply \emph{object} $\var{w}$ is tuple $(\func{pos}, r)$ where
  $\func{pos}(t) : \mathbb{R}_{\geq 0} \rightarrow \mathbb{R}^3$ is the position function over time $t$ and $r \geq 0$ is a fixed radius.
  We will write $\var{w}.\func{pos}$ to refer to the object's position function and $\var{w}.r$ to its radius.
\end{definition}

At any time, a world object occupies some region of 3D space, called an occupancy region.
The occupancy region is all points within a cube with faces at a fixed distance $r$ from a center point of $\func{pos}(t)$, as defined by the the occupancy region function.
\begin{definition}[\textbf{Occupancy Region}]
  An object's \emph{occupancy region} at a fixed time is the set of 3D space the object takes up at that time. This set
  is defined by $\func{occ}(\var{w}, t) = \{x \in \mathbb{R}^3 : \norm{\var{w}.\func{pos}(t) - x}_\infty \leq \var{w}.r \}$.
\end{definition}

When two objects have an occupancy region that overlaps, a collision is said to have occurred.
\begin{definition}[\textbf{Collision}]
  A \emph{collision} occurs when two objects $\var{w}$ and $\var{v}$ have an overlapping occupancy region at the same time $t$, called
  the time of the collision.
  This happens when $\func{occ}(\var{w}, t) \cap \func{occ}(\var{v}, t) \neq \emptyset$.
\end{definition}

A collision can be indicated by providing a collision witness.
\begin{definition}[\textbf{Collision Witness}]
  A \emph{collision witness}, or simply \emph{witness}, is a 3-tuple $(\var{w}, \var{v}, t)$ consisting of a pair of colliding objects and the time of the collision.
  Two witnesses $(\var{w}, \var{v}, t)$ and $(\var{w}', \var{v}', t')$ can be compared in time by comparing $t$ and $t'$.
\end{definition}

The continuous $n$-to-$n$ collision detection problem is to check if any two objects among a set of objects have a collision within a fixed time bound.
\begin{definition}[\textbf{Continuous $\bm{n}$-to-$\bm{n}$ Collision Detection Problem}]
  Given a time bound $T$ and $n$ world objects $\var{w}_1 \ldots \var{w}_n$, the \emph{continuous $n$-to-$n$ collision detection problem} is to find
  a minimum-time collision witness, or to prove a collision cannot occur between any of the objects within the time bound.
\end{definition}

While the general collision detection problem is defined in continuous time, it is often easier to check for collisions at multiples of a discrete time step.
\begin{definition}[\textbf{Discrete-Time $\bm{n}$-to-$\bm{n}$ Collision Detection Problem}] \label{def:problem}
  The \emph{discrete-time $n$-to-$n$ collision detection problem} is the same as the continuous-time version, except the
  collision times considered must be an integer multiple of a given time step $\delta$.
\end{definition}

The downside of the discrete-time version is that in a dynamics simulation,
objects can pass through each other (tunneling), especially if the time step is large.
For this reason, care must be taken to select a time step that is sufficiently small, so the effects of tunneling are minimized.
We will come back to the tunneling problem in the context of our proposed algorithm after we describe it.
For the rest of this paper, we will consider the discrete-time $n$-to-$n$ collision detection problem,
which we refer to simply as the collision detection problem.

\section{Existing Collision Detection Algorithms}
\label{sec:collision_detection}

We review two approaches for solving the collision detection problem from Definition~\ref{def:problem}.
In particular, we present the expected efficiency of the algorithms in terms of the number of objects $n$, the time bound $T$, and the time step $\delta$.

\subsection{Brute Force Collision Detection}

The simplest algorithm, shown in Algorithm~\ref{alg:brute}, is a brute force checking approach, where for each multiple of the time step $\delta$ from $0$ to the time bound $T$,
every object is checked for collision with every other object.

\begin{algorithm}[t]
  \caption{Brute Force Method - $\mathcal{O}(\frac{T}{\delta} n^2)$}\label{alg:brute}
  \begin{algorithmic}[1]
\Require{$\var{w}_1 \ldots \var{w}_n$, $T$, $\delta$} 
\Ensure{First collision witness $(\var{w}, \var{v}, t)$ or \textsf{None}}
%
%\State $t\gets 0$
\For{$t$ in $0$ to $T$ step $\delta$} \label{line:time}
\For{$i$ in $1$ to $n$} \label{line:fori}
\For{$j$ in $i$ to $n$} \label{line:forj}
\If{$\func{occ}(\var{w}_i, t) \cap \func{occ}(\var{w}_j, t)$}
  \State \Return {$(\var{w}_i, \var{w}_j, t)$}
\EndIf
\EndFor
\EndFor
\EndFor
\State \Return {\textsf{None}}
  \end{algorithmic}
\end{algorithm}

Although straightforward, this algorithm has two drawbacks.
First, due to the two \texttt{for} loops on lines~\ref{line:fori} and~\ref{line:forj}, it scales quadratically with
the number of objects $n$ at each time instant.
Second, due to the \texttt{for} loop on line~\ref{line:time}, the runtime grows linearly as a multiple of the number of time steps.
The scalability of this algorithm is therefore $\mathcal{O}(\frac{T}{\delta} n^2)$.
Improvements to collision detection will strive to address these two problems.

\subsection{AABB Trees}

Axis-Aligned Bounding Box (AABB) trees~\cite{aabb} are a type of bounding volume hierarchy used to efficiently solve the collision detection problem.
Bounded volume hierarchies are trees where the leaf nodes represent volumes of individual objects, and inner nodes of the tree correspond to sets
that contain every object and set below them in the tree.
An AABB tree is a binary tree, where each object and set is represented by an axis-aligned (non-rotated) bounding box.
Thus, the root of the tree will be a bounding box that contains every object, and the parent of two leaf nodes will simply be the
bounding box containing the two objects.
AABB trees can also maintain balance as new objects are inserted and updated using surface area heuristics, so that query operations remain efficient.
We do not plan to review all the details of AABB trees here, although detailed introductions are
available elsewhere\footnote{A simple introduction is available at \url{www.azurefromthetrenches.com/introductory-guide-to-aabb-tree-collision-detection}.}.

The three operations on AABB trees we will use are:

\begin{itemize}
\item \texttt{insert} - AABB trees store a world object and an associated box, usually the object's occupancy region. Insert operations take such pairs and add them to the AABB tree, possibly
  performing tree rotations to maintain balance for efficient queries.
\item \texttt{query} - Tree queries check if a given box intersects with any previously-inserted box in the tree, returning a list of colliding objects.
  Queries are performed by starting at the tree root and recursively checking if the query box intersects the
  left or right child. In the ideal case, half the objects can be discarded at each layer, leading to an efficient $\mathcal{O}(\log(n))$ lookup time. This
  may not be possible if the query box is large or the tree balance is poor.
  \item \texttt{update} - Objects in an AABB tree can have their bounds updated within an existing tree. This can prevent the need to construct
  a new tree at every time step.
\end{itemize}

\begin{algorithm}[t]
  \caption{Basic AABB Method - $\mathcal{O}(\frac{T}{\delta} n \log(n))$}\label{alg:aabb}
  \begin{algorithmic}[1]
\Require{$\var{w}_1 \ldots \var{w}_n$, $T$, $\delta$} 
\Ensure{First collision $(\var{w}, \var{v}, t)$ or \textsf{None}}
\For{$t$ in $0$ to $T$ step $\delta$}
\State $\texttt{tree} \gets \texttt{AABBTree()}$ \Comment{Creates empty AABB tree}
\For{$i$ in $1$ to $n$} \label{line:aabb_time}
\State $\texttt{box} = \func{occ}(\var{w}_i, t)$
\State $\ell = \texttt{tree.query(box)}$ \Comment{\texttt{query} returns a list}
\If{$\ell$ is not empty}
\State \Return {$(\ell[0], \var{w}_i, t)$}
\EndIf
\State $\texttt{tree.insert(} \var{w}_i \texttt{, box)}$

\EndFor
\EndFor
\State \Return {\textsf{None}}
  \end{algorithmic}
  \label{alg:aabb_basic}
\end{algorithm}

The tree balancing and query algorithm improve efficiency for spatial lookups at a single point in time, which solves the first problem mentioned
above with brute force collision detection (quadratic scalability in terms of $n$).
The update property can sometimes to be used to help with the second problem, the linear runtime with the time bound, by
heuristically inserting bloated bounding boxes
and only modifying the tree when the object moves out of its bloated box.
This will be discussed more in related work in Section~\ref{sec:related}.
%, which will be discussed
%after the basic algorithm is presented.

An AABB tree is used for broad phase collision detection, and so each object should have an associated 3D bounding box in space, at each point in time.
To get this box, we use $\func{occ}(\var{w}, t)$.
The basic AABB collision detection algorithm is provided in Algorithm~\ref{alg:aabb}.

As long as the AABB tree remains sufficiently balanced, efficient AABB tree insert and query operations make
the runtime $\mathcal{O}(\frac{T}{\delta} n \log(n))$.
This improves upon the brute force method's quadratic scalability in terms of $n$, but still has a linear dependence on
the number of time steps $\frac{T}{\delta}$.

%Heuristic methods have been proposed to improve this, or at least, reduce the linear constant associated with $\frac{T}{\delta}$.
%
%Generally, the idea is to construct a single AABB tree and perform an \texttt{update} operation at each time step to $\func{occ}(\var{w}, t)$.
%
%However, a slightly larger version of the bounding box is used instead of exactly $\func{occ}(\var{w}, t)$, so that $\func{occ}(\var{w}, t)$ can
%remain within the larger bounding box for multiple steps.
%
%This prevents the need to update the tree structure and perform tree queries at every time step for every object, but it can
%introduce false positives when tree queries detect an overlap,
%so a second phase of exact checking is added to assure the same result is obtained as
%before\footnote{This can be seen as a sort of narrow phase within the broad phase.}.
%
%Common heuristics are to bloat the bounding box by some percentage, or to use object velocity information to predict where the bounding box
%is likely to move to in subsequent steps and bloat towards that direction by some amount.
%
%These methods have parameters that need to be tuned to the situation, such as the amount of bloating.
%
%Further, their effectiveness depends strongly on the number and density of
%objects, the speed at which objects move, and the predictability of each object's motion.
%
%If the parameters are chosen incorrectly, for example bloating is too excessive, the method may perform even slower than the basic AABB approach.

\section{Collision Detection with 4D AABB Trees}
\label{sec:AABB_4D}

We propose the use of 4D AABB trees for efficiently solving the collision detection problem.
Like traditional AABB trees, 4D AABB trees include the usual three space dimensions, but they also have an additional time dimension.
Collisions are detected when two objects overlap in both space and time.
The 4D nature of the tree allows time to be tracked per-object, so that variable time steps, computed on a per-object basis, can be performed.
%
%The advantage over plain AABB trees is better scalability without situation-specific parameters and tuning.

\begin{algorithm}[t]
  \caption{4D AABB Tree Collision Detection}
  \begin{algorithmic}[1]
\Require{$\var{w}_1 \ldots \var{w}_n$, $T$, $\delta$} 
\Ensure{First collision $(\var{w}, \var{v}, t)$ or \textsf{None}}
\State $\texttt{tree} \gets \texttt{AABBTree()}$ \Comment{Creates empty AABB tree}
\State $\ell \gets$ \texttt{initializeTree(tree, }$\var{w}_1 \ldots \var{w}_n$\texttt{)}
\If {$\ell$ is not \textsf{None}}
 \State \Return {$(\ell[0], \ell[1], 0)$}
\EndIf
  \While{\texttt{true}} \label{line:while}
  \State $\var{v} \gets$ \texttt{getSmallestMaxTimeObject(tree)}
  \If{$\var{v}.t_{\max} \geq T$}
  \State \textbf{break} \label{line:break}
  \EndIf
  \State $\texttt{advanceTime(}\var{v}\texttt{, }T \texttt{, } \delta \texttt{)}$
  \State $\texttt{tree.update(}\var{v}\texttt{, }\func{occ-4d}\texttt{(}\var{v}\texttt{))}$ 
  \State $\var{u} = \texttt{resolveCollisions(} \var{v} \texttt{, tree, } \delta \texttt{)}$
  \If {$\var{u}$ is not \textsf{None}}
     \State \Return {$(\var{v}, \var{u}, \var{v}.t_{\min})$}
     \EndIf
  \EndWhile
\State \Return {\textsf{None}}
  \end{algorithmic}
  \label{alg:4daabb}
\end{algorithm}

A modified version of occupancy regions is needed that accepts intervals of time as an input,
and returns a box which bounds the states at all times within the time interval.
The function computing this region should be exact when the time interval is a single instant, but can otherwise provide an overapproximation. 
\begin{definition}[Interval Occupancy Region] \label{def:interval_occ_reg}
  An object's \emph{interval occupancy region} at some interval of time $\int{t} = [t_{\min}, t_{\max}]$ (with $t_{\min} \leq t_{\max}$)
  is a superset of the 3D space
  the object occupies at all times within the interval. This set is defined by the function
  $\func{occ-int}(\var{w}, \int{t}) \supseteq \{x \in \mathbb{R}^3 : ~ x \in \func{occ}(\var{w}, t) ~ \wedge ~ t \in \int{t} \}$.
\end{definition}

Interval occupancy functions have two additional properties which must hold:
\begin{itemize}
\item \textbf{Property 1}: \func{occ-int} should return the exact occupancy region when $\int{t}$ is a single instant in
  time (when $t_{\min} = t_{\max}$). Formally, $\func{occ-int}(\var{w}, [t, t]) = \func{occ}(\var{w}, t)$.
\item \textbf{Property 2}: If a smaller time interval is used as an input, the output should also be smaller or equal.
  Formally, If $\int{t}_1 \subseteq \int{t}_2$ then $\func{occ-int}(\var{w}, \int{t}_1) \subseteq \func{occ-int}(\var{w}, \int{t}_2)$.
\end{itemize}

The proposed algorithm requires tracking time separately for each world object, and so we augment the state with this information.
For each world object $\var{w}$, we add a time interval $\int{t}$ which we refer to as a whole using $\var{w}.\int{t}$, or
by directly naming to the individual time values $\var{w}.t_{\min}$ or $\var{w}.t_{\max}$.
This allows us to define a 4D occupancy region function.
\begin{definition}[4D Occupancy Region]
  An object $\var{w}$ has a \emph{4D occupancy region}, which is a 4D box constructed using the 3D space provided by its interval occupancy
  region function at its current time interval $\var{w}.\int{t}$, along with the 1D interval defined by the time dimension $\var{w}.\int{t}$.
  This box is defined by the function
  $\func{occ-4d}(\var{w}) = \func{occ-int}(\var{w}, \var{w}.\int{t}) \times \var{w}.\int{t}$, where $\times$ is the Cartesian product of two sets.
\end{definition}

The 4D AABB tree collision detection procedure is described in Algorithm~\ref{alg:4daabb}. 
Note that the while on line~\ref{line:while} advances a single object at a time,
rather than advancing all objects and then checking for collisions.
A proof that the loop terminates as well as algorithm correctness will be presented
in Section~\ref{sec:correctness}.

First, however, we detail each of the procedures used by the high-level algorithm.
The 4D AABB tree calls the auxiliary procedures \texttt{initializeTree},
\texttt{getSmallestTimeObject}, \texttt{advanceTime}, and \texttt{resolveCollisions}, which are described in Sections~\ref{ssec:proc_init_tree}~to~\ref{ssec:proc_resolve_collisions},
followed by a discussion on computing \func{occ-int} in Section~\ref{ssec:occ-int}.

\subsection{Procedure \texttt{initializeTree}}
\label{ssec:proc_init_tree}

The \texttt{initializeTree} procedure is used to initially insert all world objects into the AABB tree.
If every object has been inserted into the tree with minimum time $t_{\min} = 0$, such that no two boxes in the tree overlap, then \textsf{None} is returned.
%The post-condition of this procedure is that every object has been inserted into the tree with minimum time $t_{\min}$ for
%each object as 0, such that no two boxes in the tree overlap, and \textsf{None} is returned.
%
If this is impossible, because there are objects that initially collide, then a pair of colliding objects should be returned instead.

The simplest implementation, shown in Algorithm~\ref{alg:proc_init_tree} sets each object's time interval to be exactly 0,
and inserts the object into the tree.
\begin{algorithm}[t]
  \caption{Procedure \texttt{initializeTree}}
  \begin{algorithmic}[1]
\Require{$\texttt{tree}, \var{w}_1 \ldots \var{w}_n$} 
\Ensure{Pair of colliding objects at time 0 $(\var{w}, \var{v})$ or \textsf{None}}
\For{$i$ in $1$ to $n$}
\State $\var{w}_i.\int{t} \gets [0, 0]$
\State $\texttt{box} = \func{occ-4d}(\var{w}_i)$
\State $\ell = \texttt{tree.query(box)}$ \Comment{\texttt{query} returns a list}
\If{$\ell$ is not empty}
\State \Return {$(\var{w}_i, \ell[0])$}
\EndIf
\State $\texttt{tree.insert(} \var{w}_i \texttt{, box)}$
\EndFor
\State \Return {\textsf{None}}
  \end{algorithmic}
  \label{alg:proc_init_tree}
\end{algorithm}
It is possible to improve the efficiency of the algorithm by instead inserting objects with a larger interval of time, even up to the full time range $[0, T]$.
However, increasing the initial time interval could result in the detection of overlapping 4D regions, and a need to reduce time bounds of
some of the objects to eliminate the overlap.
In addition to this extra complexity, the initial performance is a one-time cost,
so efficiency improvements are not critical to the overall performance.

\subsection{Procedure \texttt{getSmallestMaxTimeObject}}

The \texttt{getSmallestTimeObject} procedure returns the object with the smallest $t_{\max}$ of all
the objects in the AABB tree, with ties broken arbitrarily.
For efficiency, rather than iterating over all the objects in the tree, the implementation should use a priority queue
implemented with something like a binary heap.
This priority queue will need to be updated every time $t_{\max}$ is changed for any object, and whenever an object is removed from the AABB tree.
In the implementation, this can be done elegantly by overriding the AABB tree methods \texttt{insert}, \texttt{update}, and \texttt{remove} to both
update the 4D AABB tree and as well as update the object's $t_{\max}$ in the priority queue.
The entire \texttt{getSmallestMaxTimeObject} procedure, then, consists of simply
returning the object at the front of the priority queue.
For this reason, we do not include its pseudocode here.

\subsection{Procedure \texttt{advanceTime}}

\begin{algorithm}[t]
  \caption{Procedure \texttt{advanceTime}}
  \begin{algorithmic}[1]
\Require{$\var{w}, T, \delta$} 
\State $\texttt{prev\_steps} \gets (\var{w}.t_{\max} - \var{w}.t_{\min}) / \delta$
\State $\texttt{next\_steps} \gets 1$

\If{$\texttt{prev\_steps} > 0$}
    \State $\texttt{next\_steps} \gets 2 * \texttt{prev\_steps}$
\EndIf

\State $\var{w}.t_{\min} \gets \var{w}.t_{\max} + \delta$ \label{line:advance_delta}
\State $\var{w}.t_{\max} \gets \var{w}.t_{\min} + \texttt{next\_steps} * \delta$

\If{$\var{w}.t_{\max} > T$}
    \State $\var{w}.t_{\max} \gets T$
\EndIf
  \end{algorithmic}
  \label{alg:proc_advance_time}
\end{algorithm}

The \texttt{advanceTime} procedure updates the minimum time for a single object $\var{v}$ to be one time step $\delta$
beyond its previous maximum time.
This is the only place where $t_{\min}$ is changed.

There is a choice of what to use for $t_{\max}$.
In our implementation, we double the length of the object's time interval when \texttt{advanceTime} is called, up to the time bound.
The result is that if the time interval is never decreased in \texttt{resolveCollisions}, which happens when the current object's 4D box does not
intersect with any other objects, then the object will only be iterated over a logarithmic number of times
with respect to the number of time steps $\frac{T}{\delta}$, in the \texttt{while} loop on line~\ref{line:while} in the high-level algorithm.
This is in contrast to the brute force method or basic AABB tree approach, where every time step requires some processing for every object, which
results in a linear scaling with respect to the number of time steps $\frac{T}{\delta}$.
In practice, the 4D boxes may overlap when large time intervals are used, and so the actual scalability will be somewhere between linear and logarithmic,
depending on (i) the distances between world objects, (ii) how fast their position changes, and (iii) the accuracy of \func{occ-int}.
Generally speaking, objects that are far from others will use larger time intervals, and therefore require less processing.
The proposed procedure is shown in Algorithm~\ref{alg:proc_advance_time}.

\subsection{Procedure \texttt{resolveCollisions}}
\label{ssec:proc_resolve_collisions}

The \texttt{resolveCollisions} procedure takes in a single world object \var{w} whose occupancy region may intersect
with other objects in the AABB tree, and reduces
the $t_{\max}$ of the passed-in object and/or the intersecting objects, in order to eliminate the intersection.
When this function is called, only \var{w} may intersect with other objects; other pairs of objects in the tree do not intersect.
If reducing $t_{\max}$ is impossible for both objects, because their time intervals are both a single instant in time, then a collision
is detected and the colliding object is returned.
If no collision is detected, then \textsf{None} is returned and we are certain no 4D boxes in the tree intersect.

The procedure uses a \texttt{queryObject} method on the AABB tree, which is shorthand for calling \texttt{query} on the tree using the object's
interval occupancy region box, and returning a list of objects with an intersection, excluding the object passed as input to \texttt{queryObject}.
The detailed procedure is shown in Algorithm~\ref{alg:proc_resolve_collisions}.
\begin{algorithm}[t]
  \caption{Procedure \texttt{resolveCollisions}}
  \begin{algorithmic}[1]
    \Require{Object with potential collisions $\var{v}, \texttt{tree}, \delta$}
    \Ensure{Object colliding with $\var{v}$ at time $\var{v}.t_{\min}$ or \textsf{None}}
\State $\ell \gets \texttt{tree.queryObject(}\var{v}\texttt{)}$
\While{$\ell$ is not empty}
\For{\var{y} in $\ell$}

\If{$\func{occ-4d}(\var{y}) \cap \func{occ-4d}(\var{v})$}
  \State $\texttt{steps\_y} \gets (\var{y}.t_{\max} - \var{y}.t_{\min}) / \delta$
  \State $\texttt{steps\_v} \gets (\var{v}.t_{\max} - \var{v}.t_{\min}) / \delta$

  \If{$\texttt{steps\_y} = 0$ and $\texttt{steps\_v} = 0$} \label{line:if_collision}
    \State \Return {$\var{y}$} \label{line:has_collision}
  \ElsIf{$\var{y}.t_{\min} < \var{v}.t_{\min}$} \label{line:increase_v}
    \State $\var{y}.t_{\min} \gets \var{v}.t_{\min}$ \label{line:increase_v_t_min}
    \State $\texttt{tree.update(}\var{y}\texttt{, }\func{occ-4d}\texttt{(}\var{y}\texttt{))}$
  \ElsIf{$\texttt{steps\_v} \leq \texttt{steps\_y}$} \label{line:decrease_v}

  \State $\texttt{new\_steps} \gets \texttt{floor(steps\_y} ~/~ 2\texttt{)}$
  \State $\var{y}.t_{\max} \gets \var{y}.t_{\min} + \texttt{new\_steps} * \delta$ %\Comment{Increase $\var{v}.t_{\min}$}
  
  \State $\texttt{tree.update(}\var{y}\texttt{, }\func{occ-4d}\texttt{(}\var{y}\texttt{))}$
  \Else \label{line:decrease_w}
    \State $\texttt{new\_steps} \gets \texttt{floor(steps\_v} ~/~ 2\texttt{)}$
  \State $\var{v}.t_{\max} \gets \var{v}.t_{\min} + \texttt{new\_steps} * \delta$ %\Comment{Increase $\var{v}.t_{\min}$}
      \State $\texttt{tree.update(}\var{v}\texttt{, }\func{occ-4d}\texttt{(}\var{v}\texttt{))}$
  \EndIf

\EndIf

\EndFor
  \State $\ell \gets \texttt{tree.queryObject(}\var{v}\texttt{)}$ \label{line:requery}
  \EndWhile

  \State \Return \textsf{None}

  \end{algorithmic}
  \label{alg:proc_resolve_collisions}
\end{algorithm}

The procedure first queries to AABB tree to see if any objects overlap with \var{v}.
If this is the case, the loop of the function essentially decreases the time intervals of either \var{v} or the colliding object \var{y}
and updates the tree with the 4D boxes corresponding to the new time bounds.
At the end of each iteration of the loop, on line~\ref{line:requery}, the tree is queried again to see if all collisions with \var{v} have been resolved.
Since the time bounds keep decreasing every time through the loop, either at some point collisions will no longer exist and the loop will exit,
or the time intervals will be reduced to a single point and a collision still exists.
In the latter case, a collision actually exists, and the colliding object is returned (line~\ref{line:has_collision}).

The time interval of one of the potentially colliding objects is always decreased when the 4D boxes intersect.
There are three ways this can happen controlled by the three branches on lines~\ref{line:increase_v},~\ref{line:decrease_v}, and~\ref{line:decrease_w}.
These are, correspondingly, increasing \var{y}'s minimum time, decreasing \var{y}'s maximum time, and
decreasing \var{v}'s maximum time.

\subsection{Computing Interval Occupancy Regions}
\label{ssec:occ-int}

One difference with the proposed 4D AABB tree collision detection method is that the user must provide the \func{occ-int} from Definition~\ref{def:interval_occ_reg}, which takes in an \emph{interval} of time and provides a box containing the occupancy regions for all times within the interval.
In the other collision detection methods described earlier, the occupancy region function needed only to provide a box for a single instant in time.
This requires computing how objects move in a given time interval.

The simplest way to compute interval occupancy regions in the discrete time setting would be to loop over all time instants and compute the smallest box
that contains the occupancy region at every point in time.
Unfortunately, this strategy would make the runtime of a call to \func{occ-int} depend on the length of the interval of time passed in,
and would reduce the performance of the approach.

If a closed-form solution of the $\func{pos}(t)$ function is available, interval
analysis methods~\cite{jaulin2001applied} can be used to compute interval occupancy regions.
Interval arithmetic methods can be used to provide bounds on functions where the arguments are each intervals.
For example, a function $f(x, y) = 2x + y$ can be used with an interval arithmetic library to compute that when $x \in [1, 2]$ and $y \in [2, 4]$,
$f(x, y) \in [4, 8]$.
In our case, if we have a formula for $\func{pos}(t)$, we could provide it to an interval arithmetic library along with any time interval to produce a bound on
$pos([t_{\min}, t_{\max}])$.
Note that this approach may provide an overapproximation of the function's true minimum and maximum, due to the well known dependency problem with
interval arithmetic.
For example, directly evaluating $f(x) = x * x$ in interval arithmetic, with $x \in [-1, 1]$, gives the overapproximation $[-1, 1]$, whereas the
true bounds are $[0, 1]$.
Accuracy is improved when smaller input intervals are provided and, in most cases, in the limit the output will approach the true minimum and maximum.
Interval arithmetic evaluation scales independently of the sizes of the input intervals, and so the computation time is $\mathcal{O}(1)$.

Many times, however, for physics simulations closed-form solutions may not be available.
Instead the dynamics of the system may be expressed with ordinary differential equations (ODEs).
These can be numerically simulated using a method such as Runge Kutta to provide the value of $\func{pos}(t)$.

If the ODE describing the movement of the object is a function of a single variable, interval methods can still be used to compute the interval
occupancy region.
This is done by simulating the system for the amount of time at the beginning and start of the desired time interval to compute the minimum and maximum values of the single variable.
Note that in this case, although the movement ODE involves a single variable, a conversion from the single variable
to 3D space can be an arbitrary closed-form function using other constant variables or properties associated with each object.
For example, in our evaluation, the single variable of an orbiting object will be the true anomaly (the angular position in orbit),
which gets converted to 3D space using other, fixed, orbital elements that are unique to each object.
Interval arithmetic is used to convert from the bounds on the single variable to bounds in 3D space.

Formally, if we have $\dot{x} = f(x)$ (where $f : \mathbb{R} \rightarrow \mathbb{R}$) for some continuous Lipschitz function $f$ (in order to guarantee existence and uniqueness of solutions)
with solution $g(t)$ (that can be obtained through numerical simulation), and $\func{pos}(t) = h(g(x))$ (where $h : \mathbb{R} \rightarrow \mathbb{R}^3$),
then we can compute bounds on $\func{pos}([t_{\min}, t_{\max}])$ by
(i) using a numerical simulation to simulate $\dot{x} = f(x)$ to get the values of $g(t_{\min})$ and $g(t_{\max})$,
(ii) performing an interval evaluation of $h([g(x_{\min}), g(x_{\max})])$.

Although more complex than the closed-form solution method, if the numerical simulation time is fast, and $x_{\min}$ and $x_{\max}$ can be looked up efficiently, the interval evaluation part of the computation remains $\mathcal{O}(1)$.
Note that if the ODE is a function of multiple variables, this approach is not applicable, and more general, reachability methods~\cite{duggirala2016msc} may be necessary to provide the bounds computed by \func{occ-int}.
The reason numerical simulation is permitted for single-variable systems is that $g(t_{\min})$ and $g(t_{\max})$ bound $g(t)$ at all intermediate times.

\subsection{Towards Continuous-Time Collision Detection}
\label{ssec:continuous_time_4daabb}

The algorithm as presented solves the discrete-time version of the collision detection problem.
Since interval arithmetic can reason about occupancy regions over intervals of time,
it is also possible to solve the continuous-time version of the problem, where
the tunneling problem can be prevented.

The main modifications required are in \texttt{advanceTime}, \texttt{resolveCollisions}, and \texttt{initializeTree}.
The \texttt{advanceTime} procedure should be changed to simply double the size of the previous time interval and set
$\var{w}.t_{\min}$ to be exactly equal to the previous value of $\var{w}.t_{\max}$.
Similarly, in \texttt{resolveCollisions}, the continuous time interval for intersecting objects should be reduced by half.
In order to terminate when objects actually collide, \texttt{resolveCollisions} should be modified to only reduce time intervals up
to a minimum time step $\epsilon$ before returning the two colliding objects.
The \texttt{initializeTree} procedure is then modified to insert objects with an initial interval $[0, \epsilon]$.

If the modified algorithm completes, then we can be sure that no collisions exists in continuous time.
Otherwise, two objects with overlapping 4D boxes can be provided, where the time intervals are less than or equal
to the minimum time step $\epsilon$, which can be chosen to be arbitrarily small.
As long as the overapproximation error provided by $\func{occ-int}$ goes to zero as the input time interval decreases,
which is true if interval arithmetic is used,
the closeness to a real collision can be made arbitrarily small by choosing a small enough $\epsilon$.
Thus, as long as systems are robustly safe, where some finite amount of space separates
interval occupation regions of all objects at all points
in time, there exists a minimum time step $\epsilon$ that can prove safety with the modified algorithm.
We plan to formalize the guarantees and correctness of the continuous-time version of the algorithm and error argument as part of future work.

\section{Algorithm Correctness}
\label{sec:correctness}

The 4D AABB Tree collision detection algorithm must meet the specification that
if collisions exist with the given step size $\delta$ and time bound $T$, the first-time collision will be returned,
and if there are no collisions then \textsf{None} is returned.
In order to prove this, we must both show that the algorithm terminates, as well as provide loop invariants that imply the desired specification.

\subsection{Termination}

In terms of loop termination, the top-level \texttt{while} loop in Algorithm~\ref{alg:4daabb} will terminate due to the specification of \texttt{advanceTime}, which advances the $t_{\min}$ of the passed-in object to be $\delta$ past its previous $t_{\max}$.
The minimum time of objects is never decreased.
The condition to break out of the loop is that $t_{\max}$ (which is not less than $t_{\min}$) reaches or exceeds $T$.
Since at each loop iteration at least one object will have its $t_{\min}$ increased by $\delta$, the break condition will become
true in a finite number of iterations.

%so that the worst-case number of iterations of the high-level loop is $n * \lceil \frac{T}{\delta} \rceil$.

The other loop with non-trivial termination is the one in \texttt{resolveCollisions} shown in Algorithm~\ref{alg:proc_resolve_collisions}.
Here, at each iteration through the loop, either the time interval of
$\var{v}$ is reduced, or the time interval of some colliding object $\var{y}$
is reduced (due to the three branches on lines~\ref{line:increase_v},~\ref{line:decrease_v}, and~\ref{line:decrease_w}).
Reducing a time interval will never cause more objects to intersect, due to
property 2 of $\func{occ-int}$ discussed after Definition~\ref{def:interval_occ_reg}.
For this reason, at each call to \texttt{tree.queryObject}, the set of intersecting objects returned must be a subset of the previous call.
Since the time bounds are reduced at every iteration, eventually either the colliding object $\var{y}$ will be omitted when \texttt{tree.queryObject} is called at the end of each loop iteration (an intersection of the 4D boxes no longer exists), or the time bound will become a single point for both objects, in which case the condition on line~\ref{line:if_collision} will become true and the procedure will exit.
%
%A bound on number of iterations through the loop can be computed by noting that in the worst case every object collides with $\var{v}$, and that an object's time bound is decreased by at least one step each time through the loop (the actual reduction is faster since we divide the number of steps by two).
%
%This again leads to an upper bound of $n * \lceil \frac{T}{\delta} \rceil$ iterations.

\subsection{Correctness}

In order to show correctness, there are two key loop invariants in the \texttt{while} loop in the high-level code in Algorithm~\ref{alg:4daabb}.
Let $t'$ be the smallest $t_{\max}$ of all the objects at the beginning of each iteration,
so that for every object $\var{z}$, $\var{z}.t_{\max} \geq t'$.
The two loop invariants are that, at each iteration of the loop:
\begin{itemize}
\item \textbf{Loop Invariant 1:} Every object $\var{z}$ has $\var{z}.t_{\min} \leq t' + \delta$.
\item \textbf{Loop Invariant 2:} There is no actual collision up to and including time $t'$.
  \end{itemize}

Upon first entering the loop, the specification of \texttt{initializeTree} guarantees that all objects are in the tree with $t_{\min} = 0$ and there are no
collisions.
The first loop invariant condition is thus initially true, since for every object $\var{z}$, $\var{z}.t_{\min} = 0 \leq t' + \delta$ ($t'$ is at least 0).
For the second condition, since the time interval for every object includes $[0, t']$, there are no collisions on this time interval, otherwise they would have been
detected by \texttt{initializeTree}.

Within the iteration of the loop, we must show that, assuming the two loop invariant conditions are true at the start of a loop iteration,
they will remain true at the end of the loop iteration.
In the \texttt{while} loop in Algorithm~\ref{alg:4daabb}, notice that
$\var{v}$ initially has $\var{v}.t_{\max}$ equal to $t'$ by the specification of
\texttt{getSmallestMaxTimeObject}.

The first loop invariant condition remains true because, in every place that
any object's $t_{\min}$ is increased, it is set to $t' + \delta$.
In \texttt{advanceTime}, $\var{v}.t_{\min}$ is assigned to the previous $\var{v}.t_{\max} + \delta$,
which is equal to $t' + \delta$.
In \texttt{resolveCollisions} on line~\ref{line:increase_v_t_min}, another object \var{y} has its $t_{\min}$ is set to $\var{v}.t_{\min}$,
which has been set to $t' + \delta$ by the earlier call to \texttt{advanceTime}.
These are the only two places where $t_{\min}$ can be increased, and so the first loop invariant condition holds.

For the second loop invariant condition, we consider two cases: either
there are multiple objects at the start of the loop iteration with $t_{\max}$ equal to $t'$, or 
\var{v} is the only one.
In the first case, the second loop invariant condition after one iteration
executes is exactly the same as before the iteration executes ($t'$ remains the same), and so the condition is trivially true.
For the second case, call $t''$ the value of $t'$ at the next iteration, such that we need to show that there were no actual collisions for any time $t$,
with $t' < t \leq t''$.
In our discrete time step setting, this is the same as $t' + \delta \leq t \leq t''$.
Since we have already proven the first loop invariant,
at the end of each loop iteration we know that
every object $\var{z}$ has $\var{z}.t_{\min} \leq t' + \delta$.
By the definition of $t''$, we further know that at the end of each loop iteration,
every object $\var{z}$ has $\var{z}.t_{\max} \geq t''$
Finally, the specification of \texttt{resolveCollisions} only allows it
to exit without a collision when the 4D AABB tree has no intersections.
Thus, there are no collisions in the time range $t' + \delta \leq t \leq t''$,
and we have proven the second loop invariant condition.

With the loop invariants available, it is straightforward to see that the loop \texttt{break} statement on line~\ref{line:break} is only executed once $t'$ reaches $T$.
From the second loop invariant, we know therefore that there
are no collisions up to time $T$, and it is correct for the collision detection method to return \textsf{None}.

Finally, we must show that when a collision exists, the one at the minimum time
is returned (there may be multiple collisions at different times).
Within the iteration of the loop, $\var{v}$, the object with the smallest $t_{\max} = t'$ has its time interval increased so that its new $t_{\min}$ is set to $t' + \delta$.
Upon calling \texttt{resolveCollisions}, if there is a collision found, it must involve $\var{v}$.
Further, since \texttt{resolveCollisions} will reduce the time interval to a single instant before returning, and since $\var{v}.t_{\min}$ is not increased in this procedure, if \texttt{resolveCollisions} finds a collision, it will be at time
$\var{v}.t_{\min} = t' + \delta$.
This, along with the second loop invariant that there are no collisions up to time $t'$,  proves that if a collision is found, it will be at the first collision time.

\section{Decomposition of the AABB Tree State Space}
\label{sec:band_splitting}

The runtime of the algorithm described in the earlier sections depends strongly on the number of objects, $n$, in the collision detection problem.
If objects stay sufficiently far away from each other, the algorithm will increase the length of the time intervals exponentially,
and has a best-case runtime of $\mathcal{O}(\log \big( \frac{T}{\delta} \big) n \log(n) )$.
If the objects frequently approach each other so that collision detection must be performed at every time step,
the runtime matches the original AABB tree method at $\mathcal{O}(\frac{T}{\delta} n \log(n))$.
In practice, the actual runtime will likely be between these two bounds, depending on
how closely objects are clustered.

Since $n$ is a critical component of the algorithm's performance, we can improve performance by trying to decomposing the
problem into independent collision detection subproblems, each with a smaller $n$.
When the subproblems are executed in parallel, the end-to-end runtime is reduced.

We formalize this static decomposition process using additional problem information provided through a \emph{potential-collision} graph.
A graph $G = (V, E)$ is a set of vertices $V$ and a set of edges $E \subseteq V \times V$.
\begin{definition}[\textbf{Potential-Collision Graph}]
  A \emph{potential-collision graph} is graph $G = (V, E)$ where each vertex $v \in V$ is associated with a world object, and an edge $e \in E$ exists between two vertices if a collision is possible between the two associated world objects.
  If no edge is present between the vertices, then a collision between the associated objects is impossible.
\end{definition}

It is possible to generate potential-collision graphs in many domains that use collision detection.
For instance, in aerospace applications altitude deconfliction is often used, where aircraft are assigned altitude bands and
aircraft in different bands will never collide.
In our evaluation, we look at orbital debris applications, where different classes of orbits can be used to create the potential-collision graph.

Once a potential-collision graph is defined, the process of splitting a single collision detection problem into multiple independent collision
detection problems can be done through graph edge partitioning~\cite{gonzalez2012powergraph,bourse2014balanced}.
Edge partitioning involves splitting a graph into disjoint subgraphs by performing cuts along vertices, where vertices that are cut get
duplicated in both subgraphs.
This is in contrast to the more usual graph vertex partitioning, which performs cuts along edges.
Independent collision detection problems can then be solved using objects associated with the vertices in each partition.
The correctness of this approach is based on the observation that a collision in the original problem must occur along an edge defined in the potential-collision graph, and edge partitioning does not delete edges.

\begin{figure}[t]
  \centering
  \includegraphics[width=0.7\columnwidth]{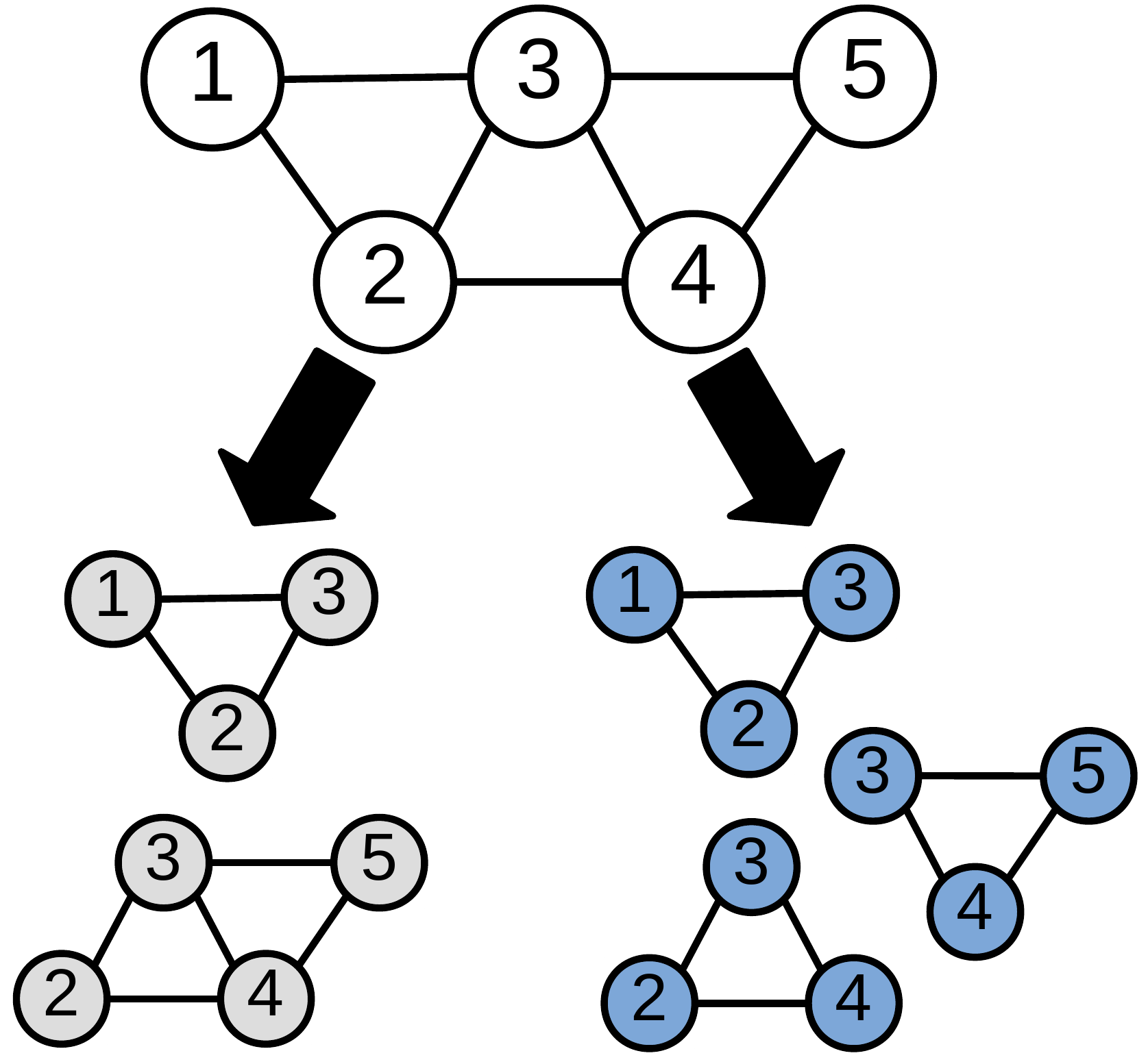}
\caption{The potential-collision graph (white, top) has $5$ nodes, whereas decomposition using edge partitioning into two partitions (grey, lower left) reduces the max node count to $4$, and a decomposition into three partitions (blue, lower right) reduces this further to $3$. }
\label{fig:potential_col_graph}
\end{figure}

An example potential-collision graph and two edge partitionings are given in Figure~\ref{fig:potential_col_graph}.
Notice that vertices corresponding to objects that have collisions in multiple partitions may need to duplicated, such as vertex 3 in the figure.

The number of partitions $p$ can be selected based on criteria such as the number of cores available.
Notice, however, that with edge partitioning we expect diminishing returns as we increase $p$, due to nodes being replicated across partitions.

\section{Space Debris Application}
\label{sec:space_debris}

We evaluate the proposed collision detection approach on a space debris collision detection application.
The U.S. Space Surveillance Network tracks around 23000 objects larger than 10 cm in orbit around Earth,
although it is estimated there are hundreds of thousands of objects between 1 cm and 10 cm,
and possibly millions smaller than 1cm~\cite{Hildreth2014}.
Due to the high velocities involved (an object in low-earth orbit moves at $7800$ m/s or about $28000$ km/hr), even collisions with small objects
can cause catastrophic damage, and further magnify the problem by creating additional space debris.
In February 2009, the Iridium 33 communications satellite collided with the defunct Russian military satellite Cosmos-2251, creating
roughly 2100 new pieces of debris larger than 10 cm~\cite{kelso2009analysis}.
Although small amounts of atmospheric drag may eventually deorbit objects so they burn up in the Earth's atmosphere, this process
can take dozens to hundreds of years, depending on the orbit.
A further concern, popularized as ``Kesler Syndrome''~\cite{kessler2010}, is that debris-creating spacecraft collisions could cascade,
resulting in an exponential increase in the amount of space debris, and threatening space access.
In response to increased launches and interest in space based services, the White House released Space Policy Directive-3,
National Space Traffic Management Policy~\cite{SpacePolicyDirective3}, which discusses collision detection and avoidance extensively
saying: ``Timely warning of potential collisions is essential to preserving the safety of space activities for all.''

Space debris collision detection is an on-going problem with real-time requirements.
The prediction speed must exceed the time needed to run the computation, in order to be able to predict collisions and warn
satellite operators to make orbital adjustments.

\subsection{Problem Setup}

\begin{figure} 
    \centering
  \subfloat[The spacecraft \emph{object} defined by $(\func{pos}, r)$.\label{fig:initialization}]{%
       \includegraphics[width=0.4\linewidth]{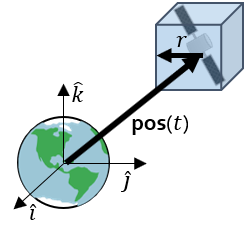}}
    \hfill
  \subfloat[The collision of two spacecraft objects $\func{occ}(\var{w}_1, t) \cap \func{occ}(\var{w}_2, t) \neq \emptyset$, where the red box shows the intersection of the two satellite occupancy regions.\label{fig:collision}]{%
        \includegraphics[width=0.4\linewidth]{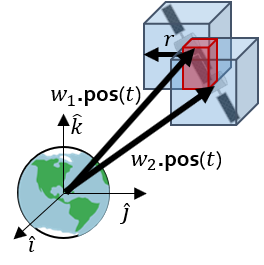}}
  \caption{Application of the \emph{object} and \emph{occupancy regions} to the orbiting spacecraft problem.}
  \label{fig:spacecraft_occ} 
\end{figure}

In this work, each spacecraft or piece of debris is an object \emph{object} $\var{w}$ tuple defined by $(\func{pos}, r)$, conceptually depicted in Fig \ref{fig:initialization}. For the space objects, $\func{pos}(t)$ is the time-dependent position of the object in three dimensional Cartesian coordinates with respect to the Earth-Centered Inertial (ECI) reference frame,
and $r$ is a representative radius of the object, which is assumed to be fixed.
There are a variety of ways to select the representative radius, including setting it to half the longest length of a satellite with known dimensions, using an estimate based on the known size class of the spacecraft,
or it may be the sum of the radii of two spacecraft~\cite{DeMars2014}.
A larger radius can be used to account for information uncertainty and model inaccuracy, but this does not affect the scalability of collision detection,
which is the main focus of this work.

The occupancy region $\func{occ}(\var{w}, t)$ of each spacecraft is a cube centered at the spacecraft position $\func{pos}(t)$,
aligned with the unit vectors of the ECI axis frame. 
The collision of two spacecraft occurs when the occupancy region of each spacecraft overlap, defined in Section~\ref{sec:preliminaries} as $\func{occ}(\var{w}_1, t) \cap \func{occ}(\var{w}_2, t) \neq \emptyset$.
These are shown in Figure~\ref{fig:spacecraft_occ}.

%The collision detection problem is to find a \emph{witness} in which any two spacecraft objects collide, as depicted in Fig \ref{fig:collision}.

%\subsection{Initialization}
A description of Kepler orbital dynamics is available in previous work~\cite{hobbs2018arch} and will only be briefly reviewed here.
An orbiting object's position and velocity can be uniquely determined from a set of six \emph{orbital elements}:
the semi-major axis $a$, eccentricity $e$, true anomaly $\nu$, inclination $i$,
right ascension of the ascending node $\Omega$, and argument of perigee $\omega$. 
When objects are only under the influence of the gravity of Earth (Kepler dynamics) only one of these parameters changes, the true anomaly $\nu$, which is
like the angular position of the object in its elliptical orbit.
The value of $\nu$ evolves according to the differential equation

\begin{equation}\label{eq:nudot}
\dot{\nu} = \sqrt{\frac{\mu}{(a(1 - e^2))^3}}(1+e\cos\nu)^2
\end{equation}
where $\mu$ is the geocentric gravitational parameter.
A nonlinear transformation consisting of three rotations involving $i$, $\Omega$ and $\omega$ then convert $\nu$ to a point in 3D space in
the ECI reference frame, where collisions can be checked.

This setup meets the requirements for computing \func{occ-int} described at the end of Section~\ref{ssec:occ-int}.
The value of differential equation of $\nu$ is a function of a single variable, and a nonlinear transformation of $\nu$ provides the position.
In the formalism of Section~\ref{ssec:occ-int}, Equation~\ref{eq:nudot} is $f$, the solution to the ODE $\nu(t)$ is $g$, and the transformation
to the ECI frame is $h$.
%
%A visual depiction of the \func{occ-int} is shown in Figure~\ref{fig:propagation}.

%\begin{figure}[t]
%\centerline{\includegraphics[width=50mm]{figures/propagation.png}}
%\caption{Propagation of the spacecraft \emph{object} \var{w} defined by $(\func{pos}, r)$, and the \emph{interval occupancy region}  $\func{occ-int}(\var{w}, \int{t})$ for $\int{t} = [t_{1}, t_{3}]$.}
%\label{fig:propagation}
%\end{figure}

\begin{figure}[t]
  \centering
\includegraphics[width=0.85\columnwidth]{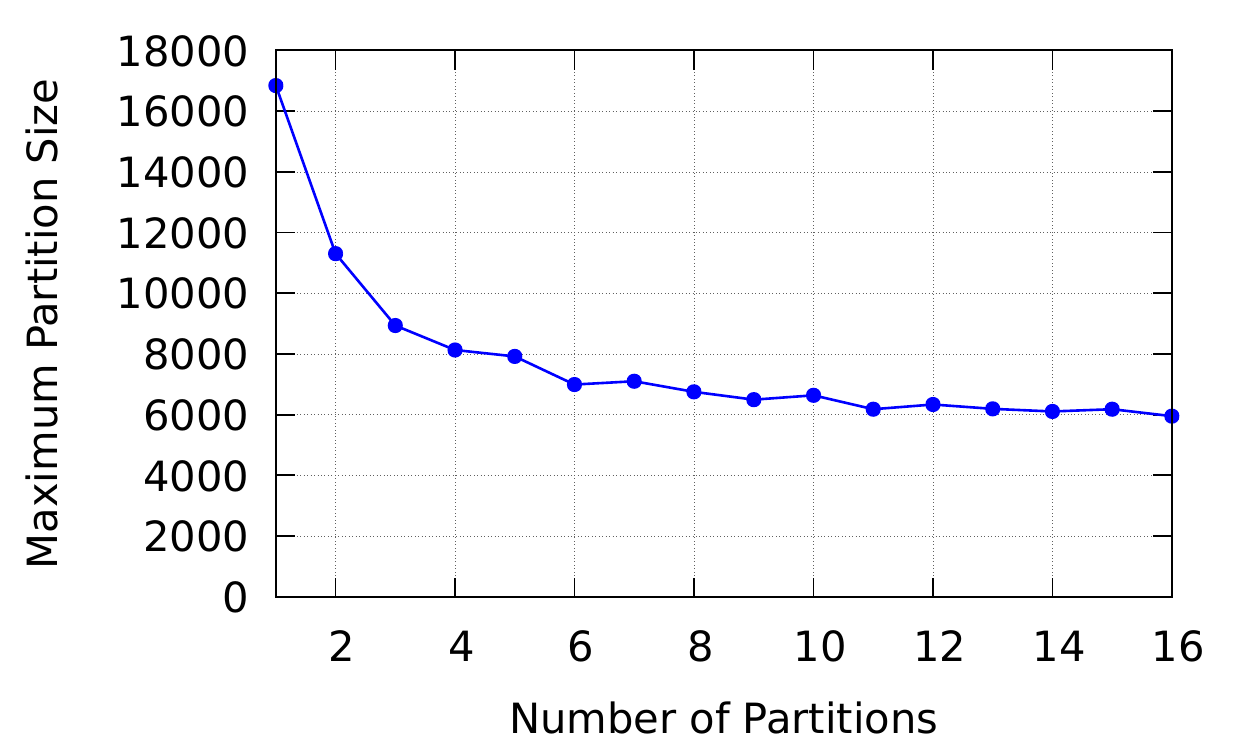}
\caption{Increasing the number of partitions generally decreases the number of objects per partition,
  although the gains are greatest when the number of partitions is small.}
\label{fig:num_bins}
\end{figure}

We also consider decomposition and parallelization of the orbital collision detection problem.
This could be done by partitioning satellites by common classes such as low earth orbit (LEO),
medium earth orbit (MEO), and geosynchronous orbits (GEO), which have well defined orbital altitude ranges.
One drawback with this simple approach is that LEO has many more objects than MEO or GEO.
Also, highly eccentric orbits (HEO) may pass through multiple altitude bands. 

These issues can both be addressed by constructing a potential-collision graph, as described in Section~\ref{sec:band_splitting}.
We use the minimum altitude (perigee) and maximum altitude (apogee) as a way to statically check if collisions are possible.
If the apogee of one object is less than the perigee of another, with small adjustments to take into account the radii of the objects, then no
collision is possible.

To perform edge partitioning, we sort the values of the semi-major axis $a$ (the average of the perigee and apogee),
and every $d^\text{th}$ value of $a$ is used to mark the edges of an altitude band, where $d$ is the total number of satellites $n$ divided by the number of
partitions $p$ ($d = \frac{n}{p}$).
%
%In other words, each altitude band is defined as the region between two spheres centered on earth, with radii equal to intervals of $a$.
%
Each altitude band defines a single partition of the potential-collision graph.
For each object, we compute its perigee as $a(1-e)$ and apogee as $a(1+e)$, and place it into all partitions that have altitudes between the two.
In this way, we attempt to balance the number of objects in each partition, in order the maximize parallelism.

\subsection{Evaluation}

We evaluate our approach using orbital elements taken from real objects, using two-line element (TLE) sets made public
by the U.S. Strategic Command on \url{www.space-track.org}.
We used the full catalog of objects larger than 10 cm taken from 3 April 2018,
which initially had $n=16840$ objects.
In order to be able to scale up or down $n$ for evaluation, when less objects were desired we simply dropped the remaining objects in the database.
When we need to evaluate with more objects, we randomly combined the orbital elements from existing objects.
Both of these approaches maintain the expected distributions of the full data set, which is clustered around low-earth orbit and not uniform across space.

Upon checking for collisions, we detected three objects in the database that seemed to be initially colliding, caused by identical TLE values.
These were two Soyuz and one resupply spacecraft docked at the International Space Station, and thus in an identical orbit.
We manually removed two of these from the database, in order to permit performance evaluation of the algorithms in the case of no collisions, making the
full catalog for our evaluation $n=16838$ objects.

We first evaluate the performance of the decomposition approach based on the potential-collision graph, when run on the full catalog ($n=16838$).
The effect of adjusting the number of partitions $p$ on the maximum number of objects in any of the partitions is shown in Figure~\ref{fig:num_bins}.
As expected, there are diminishing returns as the number of partitions increases, with the majority of the reduction obtained already by $p=6$.

\begin{figure}[t]
  \centering
  \includegraphics[width=0.85\columnwidth]{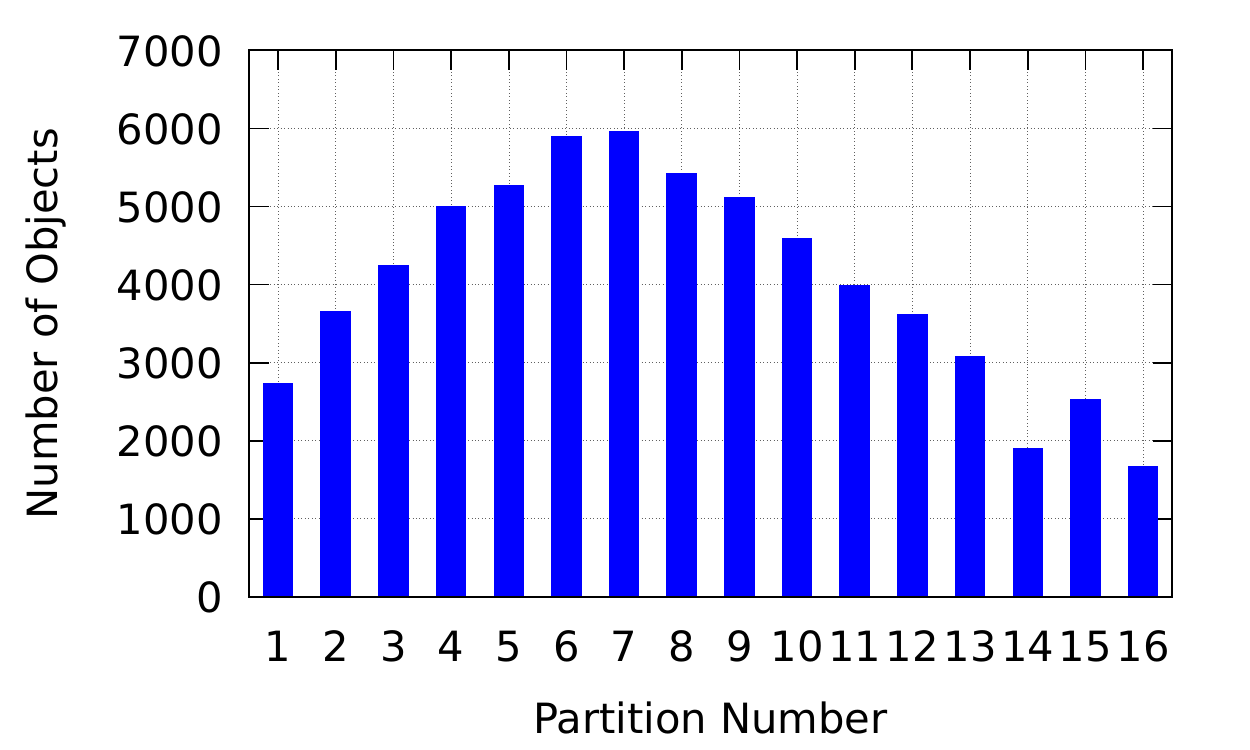}
  \caption{Edge partitioning based on semi-major axis distributes the satellites across the 16 partitions.}
\label{fig:binNumbers}
\end{figure}

Next, in the $p=16$ case, we look at how well edge partitioning was performed.
The number of satellites in each partition is shown in Figure~\ref{fig:binNumbers}.
Although generally spread out, there is still some room for improvement, as there is a range of
about $2000$ to $6000$ objects per partition.

Finally, we evaluate the overall performance of the 4D AABB method on three platforms:
an \emph{embedded} system with 1GB RAM and an Intel Atom CPU (1.33 GHz),
a \emph{laptop} computer with 16 GB RAM and an Intel i5-5300U CPU (2.30 GHz),
and a more powerful \emph{workstation} with 32 GB RAM and an Intel Xeon 8124M CPU (3.00 GHz).
All measurements were performed on Ubuntu Linux 16.04.
We also measured performance with and without partitioning.
For the partitioned versions, we set the number of partitions $p$ to be equal to the number of physical cores on the
laptop and workstation platforms, and used $p=2$ for the embedded processor, since memory was a limiting resource there.

Although applications would need to predict for hours to days of orbit time, the crucial factor
is the \emph{ratio} of the computed orbit time to the collision detection algorithm runtime.
For ease of experimentation, we fixed the orbit time to a smaller 600 seconds (10 minutes).
Since we were evaluating performance, during measurement we ensured no collisions occurred by using a sufficiently small object radius.
However, because objects in LEO move at 7800 m/s, a small time step is necessary to prevent the tunneling problem
with discrete time collision checking.
We evaluated with a time step of $\delta=10^{-4}$, so that LEO objects move about 0.78 m per step.
This is reasonable, as we envision collision boxes would be at least 10m to account for sensor errors.

\begin{figure}[t]
  \centering
  \includegraphics[width=0.98\columnwidth]{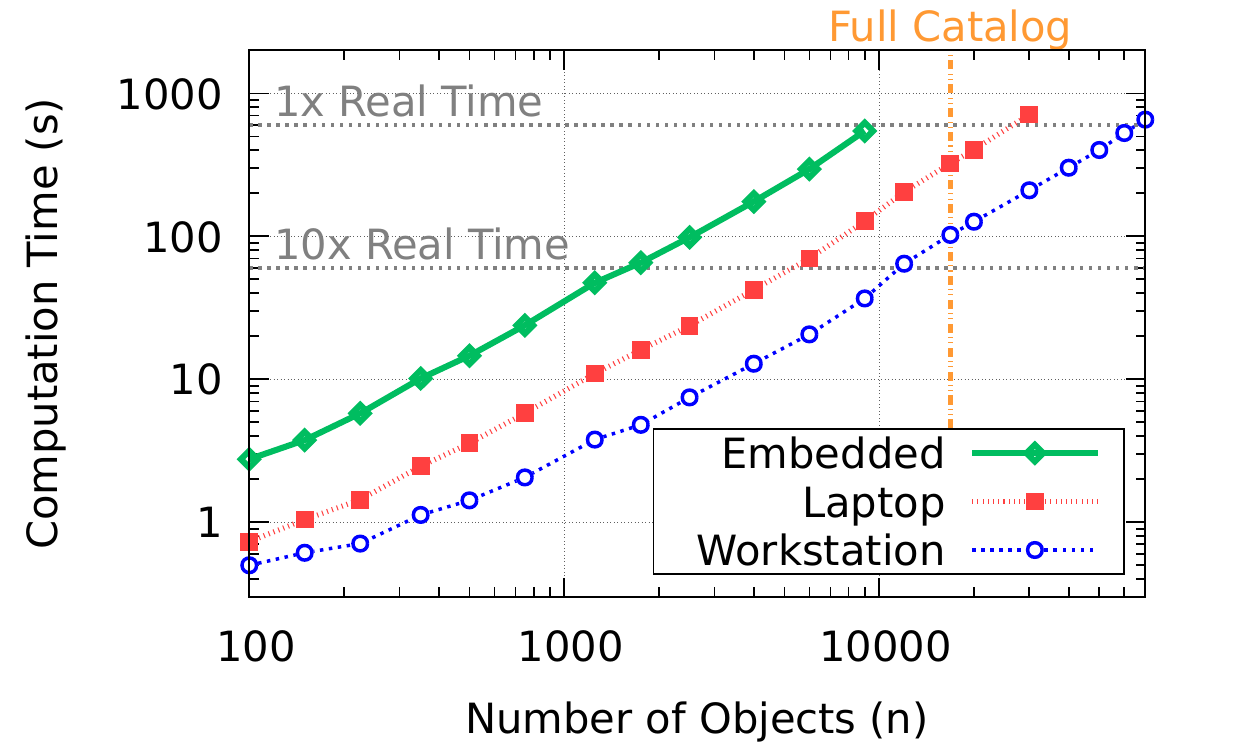}
  \caption{Runtimes for the partitioned versions on each platform from Table~\ref{tab:scalability_data}.
}
\label{fig:loglog}
\end{figure}

The full results are shown in Table~\ref{tab:scalability_data},
and a log-log plot of the data for the partitioned versions is given in Figure~\ref{fig:loglog}.
Both the laptop and workstation platforms are able to propagate the full catalog of objects and perform collision detection faster than real-time.
The partitioned workstation version is about 6x faster than real-time,
and can scale to about 65000 objects while remaining faster than real-time.
Even the embedded platform works well with smaller numbers of objects,
with $n < 1000$ being tens of times faster than real-time.
This is encouraging since embedded platforms like swarm robotics could use the approach for on-board collision maneuver prediction.

Due to the large number of steps ($10^4$ steps for each second of orbit time),
using the brute force method and even the basic AABB tree method is completely infeasible for real-time collision detection for this problem.
On the laptop platform with $n=100$, the basic AABB approach would need about 12 hours to finish checking all 10 minutes of orbit time.

\setlength{\tabcolsep}{2pt}
\begin{table}[t]
\caption{Runtime (seconds) to check 600 seconds of orbit time.}
\label{tab:scalability_data}
\centering
\setlength{\aboverulesep}{0.0pt}
\setlength{\belowrulesep}{0.0pt}
\setlength{\extrarowheight}{.2ex}
\begin{tabularx}{0.87\columnwidth}{@{}>{\columncolor[RGB]{235, 235, 235}}l@{\hskip 1.0em}lll>{\columncolor[RGB]{230, 242, 255}}l>{\columncolor[RGB]{230, 242, 255}}l>{\columncolor[RGB]{230, 242, 255}}l@{}}
%\begin{tabular}{@{}l@{\hskip 0.5em}lll>{\columncolor[RGB]{230, 242, 255}}lll@{}}
\toprule 
Objs $n$ & Embedded & Laptop & Workstation & Embedded & Laptop & Workstation \\
   & ($p=1$) & ($p=1$) & ($p=1$) & ($p=2$) & ($p=2$) & ($p=8$) \\
\midrule
100 & 3.0 & 0.7 & 0.5 & 2.8 & 0.7 & 0.5 \\
150 & 4.7 & 1.2 & 0.7 & 3.7 & 1.0 & 0.6 \\
225 & 7.4 & 1.8 & 1.1 & 5.8 & 1.4 & 0.7 \\
350 & 12.6 & 3.1 & 1.9 & 10.1 & 2.5 & 1.1 \\
500 & 19.0 & 4.5 & 2.9 & 14.6 & 3.6 & 1.4 \\
750 & 30.8 & 7.3 & 4.6 & 23.8 & 5.8 & 2.1 \\
1250 & 61.7 & 14.6 & 9.4 & 47.3 & 10.9 & 3.8 \\
1750 & 91.4 & 21.9 & 14.0 & 65.4 & 16.0 & 4.8 \\
2500 & 143.2 & 33.8 & 21.8 & 98.4 & 23.7 & 7.5 \\
4000 & 250.2 & 64.4 & 38.1 & 175.0 & 42.2 & 12.8 \\
6000 & 425.4 & 98.9 & 64.8 & 295.6 & 69.9 & 20.6 \\
9000 & 739.3 & 171.7 & 113.8 & 545.7 & 127.7 & 36.7 \\
12000 & 1185.6 & 268.2 & 178.7 & - & 204.2 & 64.3 \\
16838 & - & 422.9 & 281.1 & - & 323.4 & 102.2 \\
%20000 & - & 543.5 & 361.8 & - & 404.3 & 126.7 \\
30000 & - & 967.7 & 622.2 & - & 710.5 & 209.8 \\
%40000 & - & - & - & - & - & 301.8 \\
50000 & - & - & - & - & - & 401.9 \\
%60000 & - & - & - & - & - & 527.7 \\
70000 & - & - & - & - & - & 654.7 \\

\bottomrule
\end{tabularx}
\end{table}

\section{Related Work}
\label{sec:related}

Extensive surveys are available of collision detection methods for graphics and physics applications~\cite{jimenez20013d}.
This work falls in the category of space-time intersection methods, but rather than extruding volumes in 4D, we compute
4D bounding boxes of the space-time paths of objects.
The observation that splitting trajectories by time improves accuracy has been used before for collision detection based on
swept volumes~\cite{foisy1994safe}.

We compare our work with static interference detection methods, such as the original AABB approach~\cite{aabb}.
Modifications of the basic AABB method attempt to reuse boxes across time steps, usually
by fattening the boxes by some percentage, or using problem-specific velocity information.
These methods have parameters that need to be tuned to the situation, such as the amount of bloating, and would be unlikely to work
well for the orbit debris scenario, where velocities are high relative to the object radius.
We are similar in a sense to adaptive time-step methods~\cite{gilbert1989new}, except that we have per-object time steps.
Other than AABB trees, other data structures are available for collision detection, such as oriented bounding-box (OBB) trees~\cite{gottschalk1996obbtree}.
In this case, boxes can be arbitrarily rotated, and fast collision checking is done using the separating axis theorem~\cite{gottschalk1996separating}.
This can be advantageous since rotated boxes may have less overapproximation than axis-aligned boxes, so that tree query operations will
become more efficient.
Since we use interval arithmetic~\cite{jaulin2001applied} to reason between time steps,
AABB trees seem better suited for storing the regions of space occupied by objects in intervals of time.
Interval arithmetic methods have also been used in combination with OBB trees to provide continuous-time collision detection~\cite{redon2002fast}.
Other data structures based on sphere hierarchies have also been considered~\cite{del1992new}

A previous work considers the space debris collision avoidance problem and provides its detailed dynamics formulation~\cite{hobbs2018arch}.
There, satellites were initialized using uniform random values for their orbital elements rather than TLE sets, and only a regular 3D AABB tree
approach was evaluated, although global variable time steps were considered.
Here, using 4D AABB trees enables per-objects variable time steps, which is more efficient with large numbers of objects.

In this work, we used Kepler dynamics to propagate orbits.
Although the focus of this application was to evaluate the collision detection methods presented, more accurate methods such as SGP4~\cite{vallado2008sgp4}
also exist for orbits which take into account J2 perturbation due to the Earth not being a perfect sphere, atmospheric drag, the gravity of the Moon and Sun, solar radiation, and other effects.
Since these will modify more than one orbital element, other methods would be needed to compute \func{occ-int},
such as those based on reachability~\cite{duggirala2016msc}.
Still, even our Kepler approach could be considered as a broad-phase pass to detect potentially-colliding objects for further analysis with
the more accurate propagation methods.

\section{Conclusion}
\label{sec:conclusion}

In this work we presented a 4D AABB tree and search space decomposition approach to improve the efficiency of the collision prediction problem. 
Removing the fixed time step used in the basic AABB tree makes the 4D AABB tree more efficient because it can attempt to take large steps,
and only reduce to smaller ones when a potential collision is detected. 
We further improve efficiency by taking advantage of static problem structure to decompose the total space into a set of partitions each containing a
subset of the objects.

\bibliographystyle{abbrv}
\bibliography{hobbs,bak}

\end{document}